# A Recommender System based on the analysis of personality traits in Telegram social network

**Mohammad Javad Shayegan, Mohadese Valizadeh**

*Department of Computer Engineering, University of Science and Culture, Tehran, Iran*

*shayegan@usc.ac.ir, mersede.valizade@gmail.com*

**Abstract**

Accessing people's personality traits has always been a challenging task. On the other hand, acquiring personality traits based on behavioral data is one of the growing interest of human beings. Numerous researches showed that people spend a large amount of time on social networks and show behaviors that create some personality patterns in cyberspace. One of these social networks that have been widely welcomed in some countries, including Iran, is Telegram. The basis of this research is automatically identifying users' personalities based on their behavior on Telegram. For this purpose, messages from Telegram group users are extracted, and then the personality traits of each member according to the NEO Personality Inventory are identified. For personality analysis, the study is employed three approaches, including; Cosine Similarity, Bayes, and MLP algorithms. Finally, this study provides a recommender system that uses the Cosine similarity algorithm to explore and recommend relevant Telegram channels to members according to the extracted personalities. The results show a 65.42% satisfaction rate for the recommender system based on the proposed personality analysis.

## 1. Introduction

Social media is a social interaction place where people create information and ideas in virtual communities and networks and share them with each other. Today, people are increasingly using social media. Sharing information on the social network reflects the personality and behavior of users to some extent.

In Iran, many social networks have been successful in attracting the audience; among them, the Telegram social network has grown dramatically in attracting Iranian audiences. According to the latest statistics released by the Ministry of Information and Communications Technology of Iran, Telegram has more than 45 million active Iranian users (out of 85 million Iranian population), which includes over 60% of Iran's social network users. Moreover, the total number of posts published on Telegram per day is, on average, 2,250,000,000. Telegram is a practical social network which, as well as allowing peer-to-peer connections between users, has two features: channel and group. In a Telegram channel, the members are content consumers only, and only one person (or sometimes more) called Channel Admin is responsible for producing the content, such as audio, video, photos, and films. However, in Telegram groups, members interact and share data and can share information in the group space. Therefore, due to the information and opinions exchanging between the group members, a platform, which produces a rich data source for data analysis in this medium, is provided to analyze and check the messages of the group members.

On the other hand, personality analysis and behaviorism on social media have been the focus of many researchers in recent years. Furthermore, because of the proliferation of mobile social networks, especially Telegram, users are much more comfortable expressing their opinions in an interactive place than before. Thus this topic can be a context of analyzing users' personalities through Telegram messages. In this research, besides analyzing the individuals' personalities according to their words in Telegram groups, a recommender system is developed in the next step to recommend favorite channels to each member according to personality and behavioral analysis. For this purpose, messages from Telegram group users are collected, and then the personality traits of each member according to the famous NEO Personality Inventory[1] (NEO PI-R) are identified. For personality analysis, the study is employed three methods, including; Cosine Similarity[2], Bayes[3], and MLP algorithms[4].

## 2. Previous works

In recent researches, extensive investigations have been conducted on the analysis of users' personalities on social networks such as Facebook and Twitter. The previous studies can be divided into three general categories:

1. Analyzing users' personality on data-based social networks
2. Analyzing users' personality on image-based social networks
3. Analyzing users' personality using digital footprints

The main focus of this research is on the analysis of the user's personality in data-based social networks. Therefore, in continue, the history of data-based social networks related to this research is reviewed.

### 2.1. Analyzing the personality of users in data-based social networks
Carrie Solinger[5] began working on personality analysis on the content of Facebook users excepted that in her research, she has additionally studied variables such as the users' signed up to Facebook, the number of friends, and the chain of close friends. After the emotional segmentation of each word, she additionally created several training classes using an open-source language processing tool called Natural Language Toolkit, and the Weka tool[6]. Finally, she applied her personality analysis to the test classes. In the end, only two "Openness" and "Neuroticism" attributes obtained a high degree accuracy in analyzing users' personalities, and the rest of the attributes obtained a very low degree accuracy in the test class.

Peng[7], have worked on Facebook and have tested Facebook of Chinese users. Due to the difficulty of segmenting Chinese texts, Peng divides users' Facebook texts and contents utilizing a segmenting module called Jieba in Python programming language and calculated the number of each word occurrences using TF-IDF[8]. Afterward, he formed training and test groups based on this data and applied personality analysis. He used users' writing files and processed data using algorithms such as "Extract Properties" and "Select Proper Features," and obtained a file containing a vector-specific to each user. Finally, the SVM machine learning algorithm enabled him to predict the personality of each user.

Seidman[9] defined two main factors in addition to the 5 personality indexes as follows:

Adventitious trait: Includes observing others using Facebook and tend to communicate and learn from others on a social network

Behavioral characteristics (reflecting a person's behavior): Includes sharing a state, text, or image on a person's wall or profile

After studying these two main factors on some Facebook users, Seidman also tested five personality factors[10] of each user and recorded the results in a file specific to each one. For instance, women holding the "Openness" index are more likely to share their photos on Facebook, or women with a "Neuroticism" index tend to communicate more and have larger connections on Facebook.

In Seidman's research[9], users first take the NEO test and get their scores. Seidman also identified these two main factors for each user and performed his analysis using mean, standard deviation, variance, and linear regression. In the end, he turns out that only two Neuroticism and Agreeableness indexes are highly accurate. It should be noted that Seidman did not use any specific algorithms in his research and solely relied on two measurement factors that he introduced at the beginning of the study to meet the results of personality analysis.

Tara Marshall[11] has advanced in her research, as did Seidman, except that she has added another factor called "Response to status update" for more accurate evaluation and review. In addition to the Neo test, she provided users with a questionnaire on fields such as the amount of activity on Facebook, the using reason, the interests on Facebook, the amount of spending time of each person. Then, she measured for each person the average, standard deviation, the Cronbach's alpha coefficient, and in the end, she measured these coefficients again with five main personality indexes and recorded the results in a table. The results show that the accuracy of the "Neuroticism", "Openness" and "Agreeableness" is higher than Seidman's research.

Jennifer Golbeck[12] has used Twitter users in her research and has used further methods to offer reliable results. In her study, after collecting users' scores from the NEO test, she analyzed Twitter text and content of each user using the LIWC library[13] and re-applied the MRC[14] database on the texts and writings of users. The LIWC Library identified 81 different textual features in five groups, and the MRC identified 150,000 linguistic and psychological features. For the third step, a word-for-word analysis was performed manually on user tweets, identifying each person's score on the same person's profile. In the last step, to analyze each personality trait, she used the Weka tool and the Zero R regression algorithm.Table 1 gives a summary of some of the previous works:

Table 1: A brief comparison of previous works

| Reference No | Social network platform | Research Method | Restrictions | Year of research |
|---|---|---|---|---|
| [۱۵] | Sina Weibo | Bayes, Logistic regression algorithms | Focus on the Chinese language | 2014 |
| [۵] | Facebook | Weka Tool | Focus on Weka tool | 2014 |
| [۱۱] | Facebook | Linear regression algorithm | A low number of people in the test | 2015 |
| [۱۶] | Facebook | Decision tree, Linear regression | A low number of people in the test | 2012 |
| [۱۲] | Twitter | LIWC, Weka, Zero (R) regression | Do research only on a social network | 2011 |
| [۱۷] | Twitter | Linear regression | A low number of people in the test | 2016 |
| [۷] | Facebook | jeiba, Weka | Focus on the Chinese language | 2015 |
| [۱۸] | Twitter | Weka, Bayes | Focus on a social network | 2016 |
| [۱۹] | Instagram, Twitter, Foursquare | NUS-MSS Dataset & MBTI algorithm | The small number | 2015 |
| [۲۰] | Twitter | Link prediction & FriendTNs algorithm | Focus on a social network | 2014 |

Surveys conducted on these studies revealed that despite many kinds of research on personality analysis of social networks, any specific research had not examined personality analysis on Telegram social network because of its platform differences and certain features (channel and group). Additionally, a combination of technical works and theoretical research on personality analysis is limited in the previous works. As well, there is a lack of employing a recommender system for the personality analysis system among the literature. Therefore, this study attempts to identify users' personalities based on their behavior on Telegram and recommend Telegram channels to the users according to the extracted personalities.

### 2.2. Motivation

Analyzing users' personalities on social networks is an interesting interdisciplinary subject, particularly since a social network has many users with many contents. The popularity of Telegram, among Iranian users, a huge number of posts, the large number of its members, a large number of Telegram groups and channels, are some of the reasons why this research is being conducted specifically on Telegram. This study aims to use the users' messages of a Telegram group to analyze the personality of the group members and finally recommend Telegram channels proper to the recognized personality of people by using a recommendation framework. Recommending a few channels, close to the personality of the users, among thousands of available channels, can be very interesting to users.

## 3. Material and Methods

This research has been conducted in two phases: 1) User personality analysis and 2) Developing a recommender system to recommend Telegram channels based on the user's personality.
Figure 1 shows the process flow diagram for the research approach.

### 3.1. Phase One: Users Personality Analysis

In this section, the personality of the users in a Telegram group is analyzed after extracting and processing the messages. The following describes the stages.

#### 3.1.1 Creating a Telegram channel called Cafe K

In the beginning, a channel called "Cafe K", a channel for analysis, training, reviewing charts, and discussion about the stock market and economy, was created. After around three months by publishing valuable content, many audiences were attracted to this channel. Then, channel members were asked to join a group called "Cafe K General Group". The members could discuss in the group and left comments there. The published words through the group were an appropriate bed for analyzing people's personalities. In this group, as well as discussions on the stock market and economy issues, public and sociopolitical issues are discussed furthermore. Thus the analysis might not be considered exclusively on economic issues.

#### 3.1.2 Creating a Telegram bot

In this research, a Telegram bot was created using the C# programming language and Telegrambot library. The bot is created through Telegram bot BotFather. The bot extract message from the group according to a predefined setting.

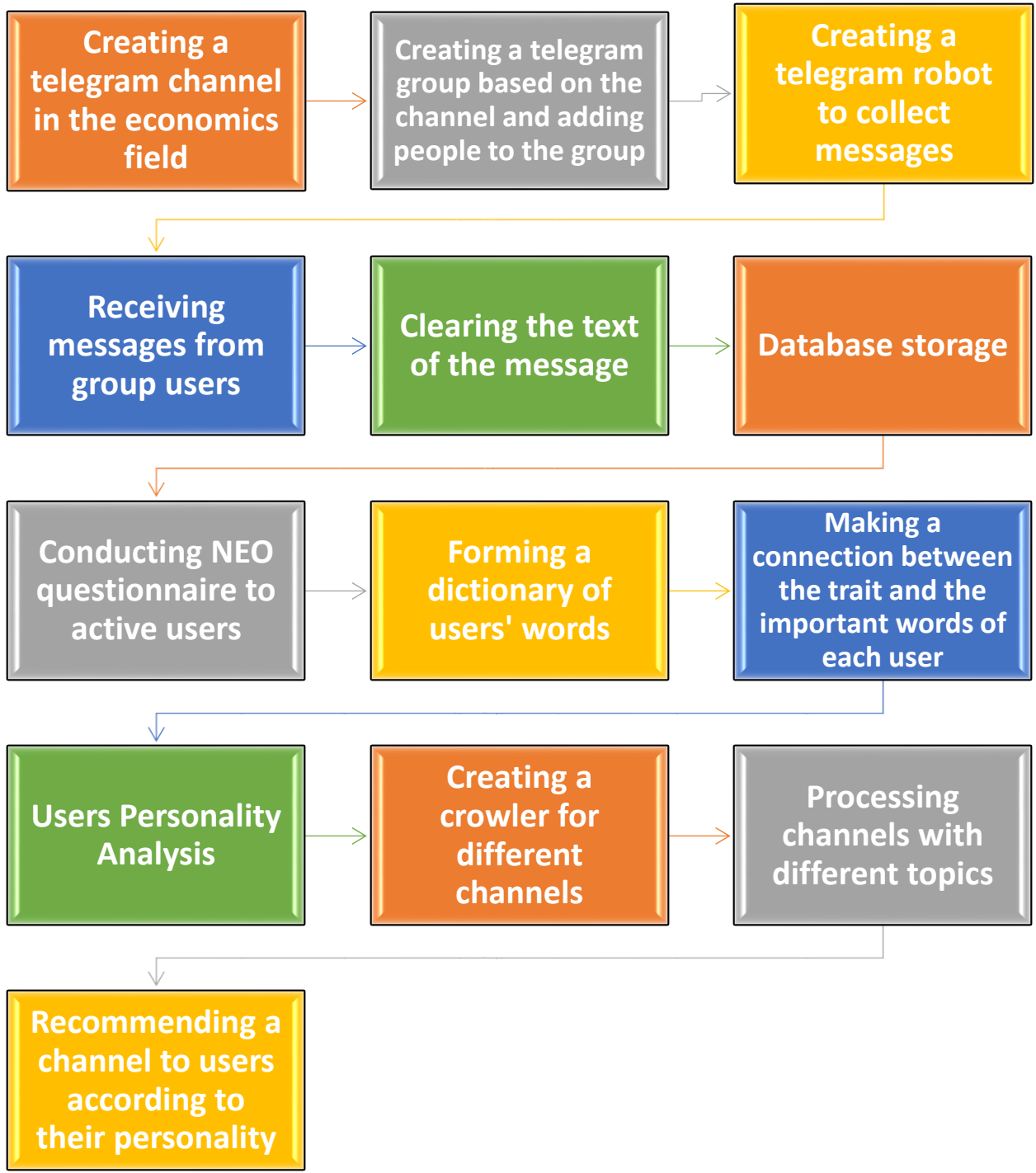

Figure 1. **Process Flow Diagram for the Research Approach**

### 3.1.3 Receiving messages from group users

The created bot is asked to listen to all the group messages and store the messages in the SQL server database. Since only the users' text messages are important, therefore in the bot code, messages including editing, games, audio, phone number, song, subtitle, film, location, photo, and empty texts are rejected and are not stored in the database.

### 3.1.4 Message Cleaning

The received users' text messages should be, without any additional characters, clean as possible by the bot. Thus, the messages are cleared, and any other characters, including dots, commas, parentheses, sharps, quotation marks, tildes, brackets, English letters, and numbers, are removed from the messages.

### 3.1.5 Database Storage

The received correct messages by the bot are stored in the SQL server database, and a table is created with the columns: name of the person who sends the message, his message, and the date and time of sending the message. Therefore, each person's messages are displayed with his ID. In the last step, using the "group by" command, a report is issued from the database to discover who has sent the most messages or talked the most in the group. After the execution of this order, it became clear that eight people in this group told more than the others. In this study, these users are referred to as so-called "active users".

### 3.1.6 Conducting a NEO questionnaire with the participation of active users

In the previous steps, the users' messages were stored in the database. Now, we have to conduct a test for personality analysis. According to the previous studiesd and online questionnaire based on the NEO-psychological test (NEO PI-R), five personality traits of each user are identified, and each user is given a score based on these five characteristics. In fact, users complete the NEO test carefully and accurately to assign a vector to each of them finally. However, in a smaller number of studies, researchers have used the Myers-Briggs (MBTI) method instead of the NEO-test [13]. We conducted a NEO questionnaire for active users. This questionnaire consists of 60 questions, which are given five traits to each person with a score for each trait. Therefore, at this step, each person has five traits, each of which has a value of 60 points. After specifying the traits of the users, the highest scored trait of each user is registered as the index trait of that user. Actually, each user is labeled with a trait that has the highest score among the five traits of the NEO-test.

### 3.1.7 Creating a dictionary of users' words

After receiving the NEO inventory of each user, the stored messages of these users in the database are processed in the Python programming language. This processing is executed to create a dictionary of the words users have used in the group. Notably, because of the frequent use of the famous LIWC dictionary in the previous works and the lack of a suitable dictionary in Persian, a dictionary has been formed somehow in Persian in this section. In such a way that the text messages of each person, which are identified with the same person's ID, are first separated word by word by the Postagger function, using an integrated tokenizer. In this function, extra letters, linear jumps, spacing, and extra items are removed. After these steps, each user has two unique files called the Usermessage and profile. In the Usermessage file, all the messages that belong to the user are stored. In the profile, the vector of each person is kept. It stores the user's messages in the form of [item, value], and the number of repetitions of that word. Thus, in the beginning, each person has a dictionary of self-used specific words. In the next step, we extract the important words of each user from this dictionary.

### 3.1.8 Establishing a relationship between each user's important trait and words

At this step, after constructing a dictionary for each group member, a global dictionary called IDF is created. It includes all the words extracted from the messages along with the number of repetitions. The IDF formula is as follows:

IDF = is the logarithm of the total number of contents dividing by the contents that contain the word.

(Formula 1)

$$\text{idf}(t, D) = \log \frac{N}{|\{d \in D : t \in d\}|}$$

In this formula, the numerator represents the total number of documents (here is the all of the Usermessage files), and the denominator represents the number of documents that contain the word X.

TF for each person indicates the number of repetitions of a particular word divided by the total number of that person's words. The TF formula is as follows:

(Formula 2)

$$tf_{i,j} = \frac{n_{i,j}}{\sum_k n_{k,j}}$$

After calculating these two variables, it is necessary to select more valuable and significant words from each user's messages. In other words, in this study, words are useful for us that have a high frequency. For this purpose, it is necessary to calculate the combination of these two variables, TF*IDF [24]. Thus, another dictionary is created to store the TF*IDF value of each word. Now we can say that each user has used which words more in the group, according to his trait (through the NEO test). For instance, a person who is suffering the Neuroticism has used the words anger, stress, depression, and fatigue more, so it can be concluded that this vocabulary is suitable and important for diagnosing people suffering the Neuroticism. It should be noted that in doing so, we will have a dictionary of different words related to different traits.

### 3.1.9 Analyzing users' personalities

In order to calculate similarities among all pairs of members, it is necessary to use similarity detection methods. The following describes two personality analysis approach employed in the current study.

### 3.1.9.1 Personality analysis using the Cosine Similarity algorithm

One of the most popular similarity detection algorithms between two variables is Cosine Similarity algorithm. There are two lists of test data and training data in this section. These two variables are given as inputs to the Cosine Similarity function, and the Cosine Similarity function returns the similarity as the output. The output of this algorithm shows the user with the most similarity to one of the active users (training users). In the same way, all similar users are extracted.

### 3.1.9.2 Personality analysis using Machine Learning

In this study, the SKLearn[21] library in the Python programming language is used, aiming to meet more efficiency than two algorithms: MLP in neural networks and the Bayes.

To analysis, after adding the SKlearn library, a matrix is formed in which every element is the multiplying result of each user's dictionary (TF*IDF) by the words' IDF. Thus, the attributes are IDFs that make the columns of this matrix in the (IDF0, IDF1, IDF2, …) formats. Then, two train and test matrices are created so that 33% of the matrix is used for test and 66% for train matrices. The machine learning algorithm should analyze 33% of its test users and return the result to the target list. It should be noted that the target list has been created in two ways to get a better result: 1) using the output of the Cosine Similarity function 2) using the NEO inventory test result of all users (manual method). After performing the algorithm, the personality analysis result is calculated as the output in the target list.

### 3.2. Phase Two: Creating a Recommender System to recommend users the channels

In this phase, after defining each user's personality, related channels to the users' personalities are recommended to them using a recommender system.

### 3.2.1 Creating a crawler for channels

Once each user's personality is acquired, it is time to recommend him Telegram channels in different fields depending on his personality. In this regard, channels with the titles: Digital Currency, Bank Era, Banker, Churchill (with contents such as stock market signals, market analysis and review of the stock exchange and economic meetings), Cinema, Programmers, Dr. Salamat, God friendship, Forough Farrokhzad, White Poetry, Anecdotes and Stories, Child Psychology, Stylish Catering, Lean Monologue, Sports 3 and the World of Economics were evaluated. For this purpose, a crawler was written in the C# programming language and HTML agility pack library. This crawler extract posts from the channel.

### 3.2.2 Telegram channel processing

In the desktop version of the Telegram application, a backup in the form of a web page was taken from the above-mentioned channels. The extracted data includes the main DIV (HTML tag) and several nested DIV. The crawler's job is to get small DIVs (or channel posts). These posts are stored in the SQL Server database under the same channel name. It means, like group members, the database includes channel names, posts, date and time. Finally, the CSV file of this data is constructed. All the steps performed for each user are also executed for each channel. This means that we considered each channel similar to a person, along with its personality. Therefore, each channel with its title has a profile that includes the TF-IDF vector within, and from this vector and using the Cosine Similarity function between the critical words of the channel and traits, proportional channels to each trait can be obtained. Furthermore, precisely according to the developed process for users, the machine learning mechanism, Bayes and MLP algorithms are used to obtain the personality of each channel that the results of which will be described in more detail below.

### 3.2.3 Recommending Telegram channels with various topics to users

Using the output of the Cosine Similarity function, the Bayes, and MLP algorithms, you can recommend channels corresponding to the user's personality. At the end of the work, each user is asked about the recommended channels, and then the results and accuracy are obtained.

## 4. Results and Discussions

Before further discussion for the findings, the following section presents an initial statistical analysis.

### 4.1 Preliminary Statistical Analysis

The examined messages of this research are the messages of active users in the group and who have the largest number of words to increase the validity of the results. Furthermore, the channels that were reviewed and evaluated to recommend to users have an average of 17,000 messages and 254,000 members.

Figure 2 shows the number of each person's messages separately, and the activity of each person in the group. As can be seen, the first user had posted 9995 messages, the largest number of group messages. Most users had less than 2000 messages. This number indicates that many people in Telegram group are quite. This common behavior causes limitations on sufficient data collection for subsequent data mining steps.

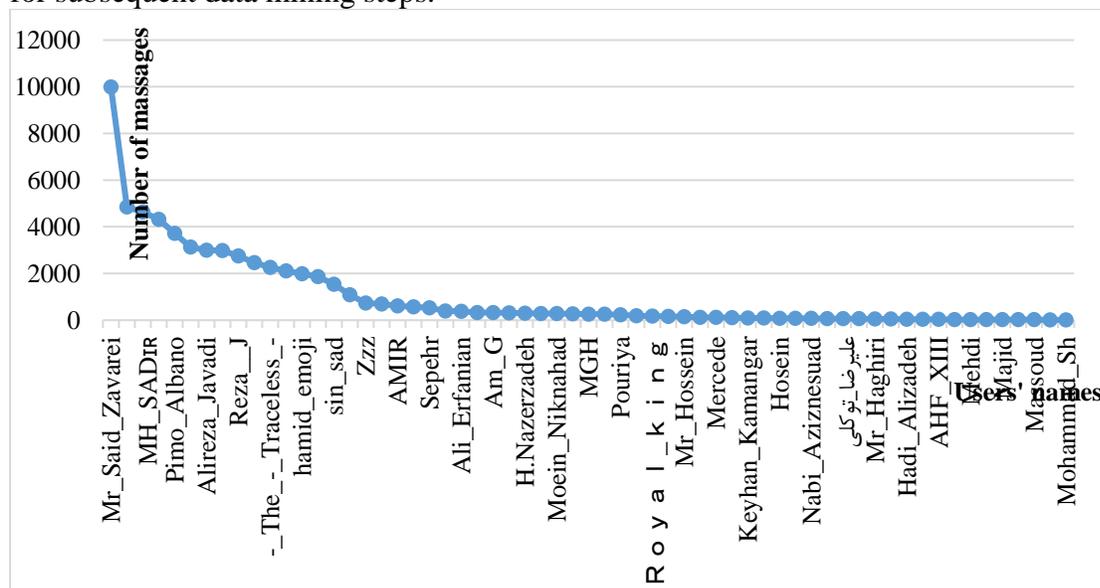

Figure 2. Number of messages of each person

## 4.2 The accuracy of the results

In this research, three methods are used to analyze the users of a Telegram group personality. Figure 3 shows the accuracy of the three algorithms. The accuracy of the Cosine Similarity algorithm was 76.25%. The results show the accuracy of the Bayes and MLP algorithms were 82.5% and 81%, respectively. The results show the Bayes has the best functionality in the current research.

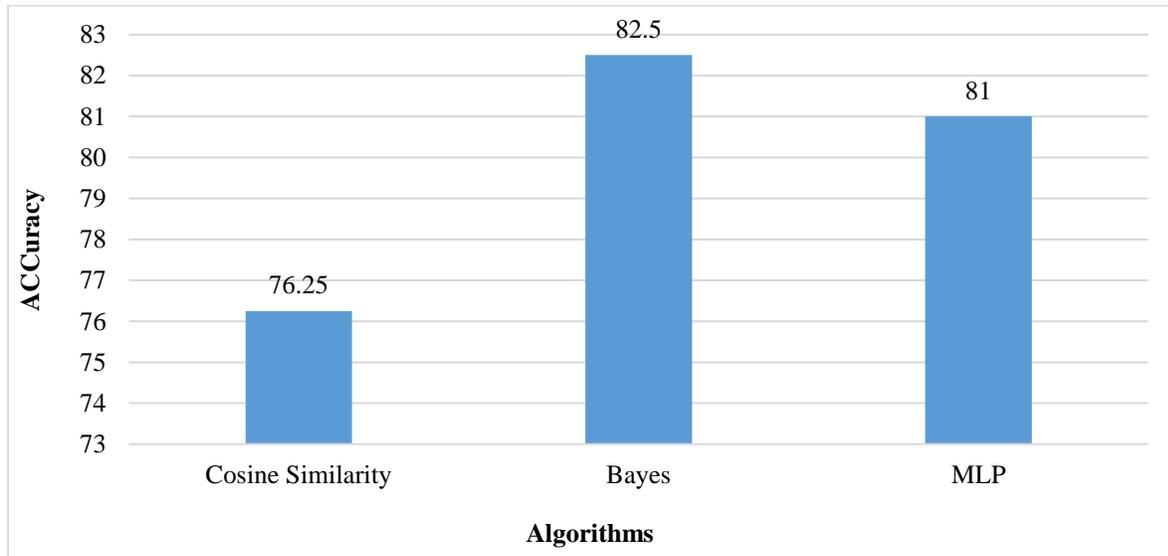

Figure 3. The accuracy of algorithms for similarities in personality analysis

Figure 4 is a sample of the output for Cosine Similarity algorithms. It shows details of measuring the similarities. For instance, a person called Behnam is similar to a person called Abbas using a Cosine Similarity algorithm; the scores of the NEO-tests are shown separately for each one. This output shows that using the set of words that Abbas used in the group and the similarity between his words and Behnam's; these two people are supposed to have similar personalities in the algorithm. Moreover, in this figure, the scores of Behnam and Abbas NEO-test are very similar, and out of five test scores, four are in the same range, which shows the ideal result. As another instance, a person called mrr is similar to a person called Mr. Said Zaverei; as can be seen, these two people have only three of the five their NEO-test traits similar in a period. A remarkable note about this figure is for some people with a blank in front of their names because no similarity has been inferred due to having no conversation in the group.

```
Users_result_analysis_Cosine_Similarity.txt - Notepad
File Edit Format View Help
Mohammad_Hassan_Torabi,39,29,40,46,51------
R_Moafimadani,26,46,42,43,54-----
Behnam,28,32,34,44,38------Abbas,36,35,40,52,44  s=4
M._Farham,23,42,42,46,50-------
A_K,33,38,39,40,47-------
Hadi_Alizadeh,29,40,36,38,45---------Abbas_Davari,28,44,39,27,37 s=5
SafaAli_Mozafarnejad,34,34,34,40,45-----Abbas,36,35,40,52,44 s=4
Reza_Sadoughi,45,45,46,46,46-------
reza,39,36,42,39,38--------
Somayeh,24,40,39,41,51-------
Arman,52,27,49,44,20-------
Mehdi,33,30,40,41,48-----Abbas,36,35,40,52,44 s=4
meyfi,43,27,38,37,37------
Mercede,31,44,40,47,53-----Mr_Said_Zavarei,26,43,39,38,46 s=4
Vahid_m,34,42,36,32,49----
Mr_Forex,31,40,34,53,41-----Mr_Said_Zavarei,26,43,39,38,46 s=4
Hamide,23,43,36,48,45-----Abbas_Davari,28,44,39,27,37 s=4
Hosein,35,34,38,45,52-----Abbas,36,35,40,52,44 s=3
Nabi_Aziznesuad,20,38,37,49,55-----
Arash,23,36,32,45,49-----Mr_Said_Zavarei,26,43,39,38,46 s=4
Mohammad_Mahdi,32,37,36,46,50------
Vara,28,32,38,43,46----Setmax,28,45,43,46,47 s=5
Mohammad_Sh,26,31,41,41,44-----
Pedram_Zahedi,34,41,43,37,42----Abbas,36,35,40,52,44 s=4
Keyhan_Kamangar,26,33,35,38,44----Abbas,36,35,40,52,44 s=4
mrr,21,36,42,45,51----> Mr_Said_Zavarei,26,43,39,38,46 s=3
Majid,34,43,43,46,53-----Setmax,28,45,43,46,47 s=4
Mr_Haghiri,33,35,36,43,50--------Mr_Said_Zavarei,26,43,39,38,46 s=4
AHF_XIII,26,41,38,39,44--------Mr_Said_Zavarei,26,43,39,38,46 s=5
61/16 /5 *100 = 76.25
```

Figure 4. The output of the Cosine similarity algorithm

### 4.3 The accuracy of the obtained results for the recommended channels

As previously mentioned, in the last part of this research, a number of the Telegram channels are recommended to users according to their personality analysis. To evaluate the recommended channels, the same approach for personality analysis of the user method was employed. Meaning, each channel with different contents is considered as a person with different personalities. Then, using this approach, each channel is mapped to the training data according to the words and phrases has used.

For the recommender system, the Cosine Similarity algorithm has been used to obtain channels proper to each person's personality. To measure the accuracy of the results, the users were asked, "How many of these channels have been proper to your taste?". Therefore, the percent of satisfied people show the accuracy of the recommender system. The results show that 65.42% of users have positive feedback on the recommended channels.

Figure 5 shows the level of satisfaction from the recommended channel for each member. As it is seen, unlike three of these people who had a negative response to the recommended channels, ten people fully agreed with the recommended channels, and the rest agreed only with some recommended channels. These results seem desirable due to the lack of empirical research in this field. It is also notable that the zero values belong to people which their NEO-test result was moderate (neither strong nor weak) for any personality trait.

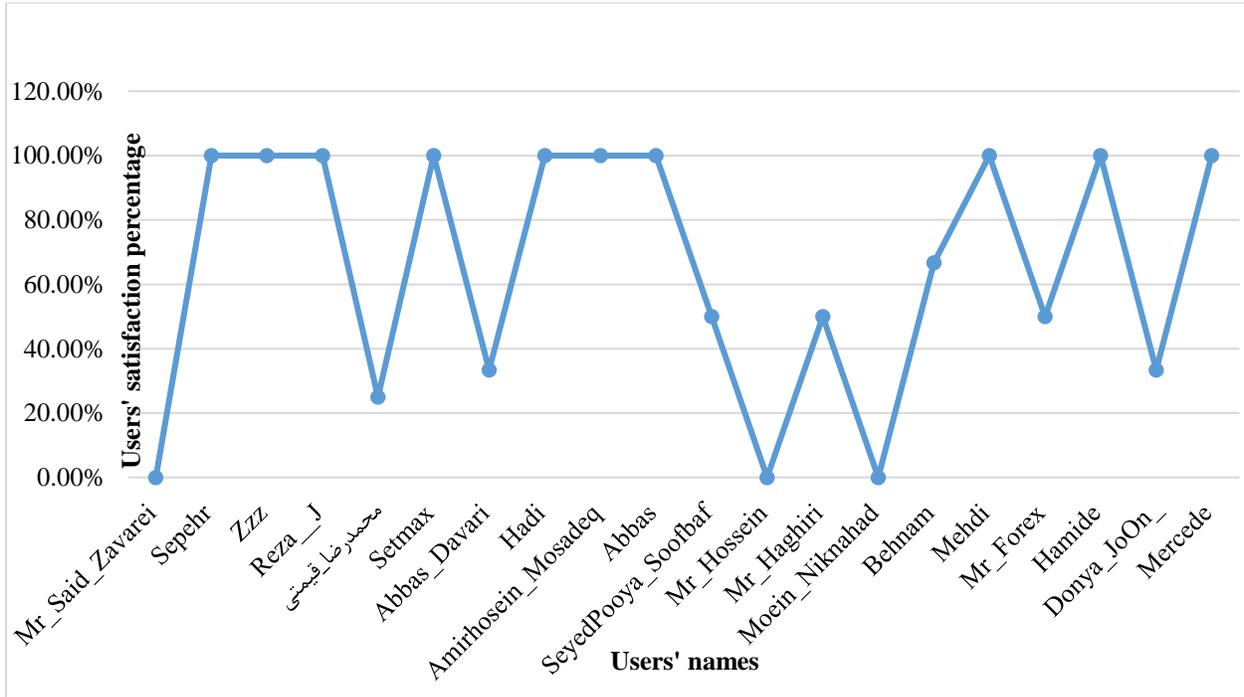

Figure 5. Each member's level of satisfaction from the recommended channel

## 4.4 Comparison

It is worth mentioning, the current study differs from previous research in terms of language, social network type, dataset, implementation, and methodology. Still, to show the proximity of the results to previous works, a brief comparison with some previous works are presented in Table 2:

Table 2: Comparison of previous works in terms of accuracy

| Reference No | Accuracy | Social network platform | Algorithms and implementation tools in each research |
|---|---|---|---|
| [22] | 70% | Sina weibo-Bayes | Logistic Regression |
| [16] | 70% | Facebook | Weka Tools, Bayes |
| [18] | 76% | Facebook | Regression |
| [21] | 54% | Facebook | Decision tree, Linear regression |
| [8] | 11%-18% | Twitter | LIWC Dic, Weka Tools, Zero (R) Regression |
| [1] | 65% | Twitter | Linear regression |
| [7] | 50%-60% | Facebook | Jeiba Tools, Weka Tools, Regression |
| [12] | 60%-70% | Twitter | Weka Tools, Bayes |

Due to the differences in the social network platforms and implementation algorithms in the previous works, we consider the average accuracy of the results, which is around 60%. Among these works, [8] has the lowest accuracy. The authors in [8] explicitly stated it as the main weakness of their work. Therefore, considering many differences between the forthcoming research and previous works in the social network platform, the users' language, the employed algorithm, the current study looks like[7]. In[7] the social network is Facebook, and the language of users is Chinese. Furthermore, the implementation approach and the employed algorithm are somewhat different. However, in the present study, the social network is Telegram, and the language is Persian. Meanwhile, we tried to collect the Persian dataset with many efforts to create a Persian word dictionary. Moreover, because of the nature of Telegram in Groups and Channels, several methods have been implemented, and several experiments have been used to obtain a more desirable and closer to reality result.

## 5. Conclusions and Suggestions

This study aims to analyze the users' personalities automatically, and then, using obtained results, design a recommender system to suggest appropriate Telegram channels to the users.

Thus, the current study had two general phases. Users' personality identification and developing a channel recommender system according to the obtained personality for the users and channels.

For the first phase, the users' messages were extracted from a Telegram group. Then, we produced a dictionary for the extracted messages. Then, a relationship between each user's trait and his words were established. At the same time, the personalities of the users were identified by conducting a NEO survey. The personality analysis was done by using Cosine Similarity, Bayes, and MLP algorithms. The accuracy of the Cosine Similarity algorithm was 76.25%. The results show the accuracy of the Bayes and MLP algorithms were 82.5% and 81%, respectively.

In the second phase, Telegram channels were analyzed as users, and the personality of each channel was evaluated. We measured the personality of a Telegram channel similar to the method was employed for a user's personality. In the last part, according to the similarity between the channels' content and the personality of the group members, channels proper to each user's personality were recommended. Followings are the features of the proposed approach:

 • Automatically extract users' personalities from their messages on a mobile social network platform such as Telegram.

 • Creating a dictionary in the Persian language for analyzing personality traits for the first time (such as the LIWC dictionary).

 • Targeting Telegram as a special mobile social network with different features such as Group and Channel.

 • Using three different algorithms to meet more accurate results.

 • Analyzing channels with different characters like users and group members.

 • Receiving positive feedback from users for the recommended channels with 65.42% satisfaction rate and finally

 • Providing a capable system to successfully identify users' personalities and recommend users' favorite channels among thousands of available channels.

The discussed topics of this research are novel issues that provide several research opportunities. Since personality analysis is a multidisciplinary field, future research can be improved in each dimension of the current research. For instance, future research enriches the dictionary section in the Persian language like the LIWC dictionary in English. Other algorithms that can be employed to assure the personality analysis results with high percent accuracies and efficiencies. This achievement will be extendable in other languages as well. Among other recommended opportunities, we can mention utilizing the individual personality traits in business areas.

**Declarations**

**Funding**: This research did not receive any specific grant from funding agencies in the public, commercial, or not-for-profit sectors**.**

**Conflicts of interest**: The authors declare that they have no conflict of interests.

**Informed consent**:  Informed consent was obtained from all individual participants included in the study.